\documentclass[]{article}
\pdfoutput=1
\usepackage{fullpage}

\usepackage[up]{subfigure}
\usepackage{amsmath, amssymb, amsthm,graphicx}
\usepackage{amsfonts}
\usepackage{color}

\newcommand{\R}{\mathbb R}

\title{An uncertainty principle underlying the pinwheel structure in the primary visual cortex}

\author{D.Barbieri\footnote{Dipartimento di Matematica Universit\`a di Bologna, Italy}, 
G.Citti \footnote{Dipartimento. di Matematica. Universit\`a di Bologna, Italy}, G.Sanguinetti \footnote{DEIS. Universit\`a di Bologna, Italy - Universidad de la Rep\'ublica, Montevideo, Uruguay.}, A.Sarti \footnote{CREA. Ecole Polytechnique. Paris. France}{\phantom A}\footnote{Authors are listed in alphabetical order.}}
\date{}

\begin{document}

\maketitle

\section{Introduction}

The visual information in V1 is processed by an array of
modules called orientation preference columns \cite{Hubel1977Ferrier}.  In some species including humans,
orientation columns are radially arranged around singular points
like the spokes of a wheel, that are called pinwheels. The pinwheel
structure has been observed first with optical imaging techniques \cite{Bonhoeffer1991Isoorientation} and more recently  by in vivo two-photon imaging proving their organization with single cell
precision \cite{Ohki2006Highly}. Many hypotheses of morphogenesis of
pinwheels during visual development have been proposed e.g. \cite{Wolf1998Spontaneous}\cite{Bressloff2003Spherical} and many reviewed in \cite{Swindale1996Development}\cite{Simpsom2009Theoretical}, but at the
present the problem is still open \cite{Huberman2008Mechanisms}. In
this research we do not intend to propose another model of pinwheel
formation, but instead provide evidence  that pinwheels are de facto
optimal distributions for coding at the best angular position and
momentum. In the last years many authors have recognized that the 
functional architecture of V1 is locally invariant with respect to the symmetry group of rotations and translations $SE(2)$ \cite{BenShahar2004Geometrical}\cite{Zweck2004Euclidean}\cite{Bressloff2001Geometric}\cite{Citti2006Cortical}\cite{Franken2009Crossing}\cite{Sarti2008Symplectic}\cite{Duits2007Image}. In the present study we show that the orientation cortical maps
measured in \cite{Bonhoeffer1991Isoorientation} to construct pinwheels, can be
modeled as coherent states, i.e. the geometric configurations best
localized both in angular position and momentum.  The theory we adopt is based on the well known uncertainty principle, first proved by Heisenberg in quantum mechanics \cite{Cohen1978Quantum}, and later
extended to many other groups of invariance \cite{Folland1989Harmonic}\cite{Carruthers1968Phase}\cite{Isham1991Coherent}.
This classical principle states an
uncertainty between the two noncommutative quantities of linear
position and linear momentum and it has been adopted in \cite{Daugman1985Uncertainty} to model receptive profiles of simple cells as minimizers defined on the retinal plane.
Here we state a corresponding principle  in the cortical geometry with 
$SE(2)$ symmetry based on the uncertainty between the two noncommutative
quantities of angular position and angular momentum. By computing its minimizers we obtain 
a model of orientation activity maps in the cortex. As
it is well known the pinwheels configuration is directly constructed
from these activity maps \cite{Bonhoeffer1991Isoorientation}, and we will be able to formally reproduce
and justify their structure, starting from the group symmetries of the functional architecture of the visual cortex.
 The primary visual cortex is then modeled as an integrated system in which the set of simple cells implements the $SE(2)$ group, the horizontal connectivity implements its Lie algebra and the pinwheels implement its minimal uncertainty states. 


\section{The functional architecture of V1}

\subsection{Simple cells as elements of a phase space}

  Receptive fields of simple cells of the area V1 are classically
 represented as Gabor-like
 filters able to accomplish an analysis both in space and frequency \cite{Daugman1985Uncertainty}\cite{Lee1996Image}\cite{Jones1987Evaluation}.
Each filter is identified by four parameters $(q_1, q_2, |p|,
\theta)$ where
 $q= (q_1, q_2)$ are coordinates in the $2$D cortical layer $C$,
 $|p|$ and $ \theta$ are the engrafted variables in the sense of Hubel \cite{Hubel1998Eye}. The coordinate
 $|p|$ codes the modulus of frequency and
 $\theta$ the orientation preference, so that they
 can be interpreted as a choice of
 polar coordinates in the frequency variables.
 Since the variables of position and frequency are constitutive
 of the
 phase space, the set of simple cells is classically
 labeled by parameters of a 4D
phase space  which represents the cotangent space $T^*(C)$ of the
2D cortical layer. We also explicitly note that Gabor
filters are complex valued, so that the cortical signals related to this 
cells are also complex.

\begin{figure}[!ht]
\centering
\subfigure[ ]{
\includegraphics[width=8cm]{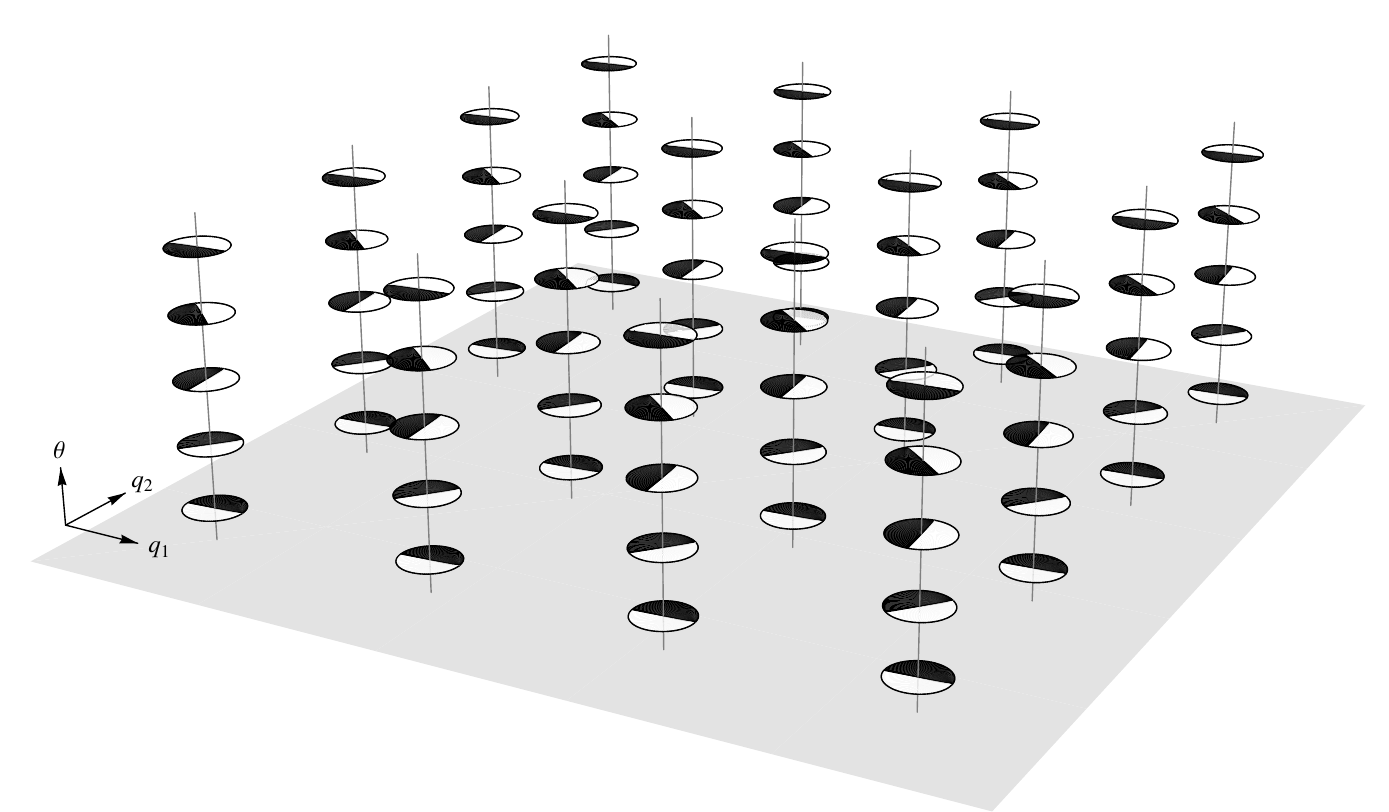}
\label{fig1}
}\\
\subfigure[ ]{
\includegraphics[height=3cm]{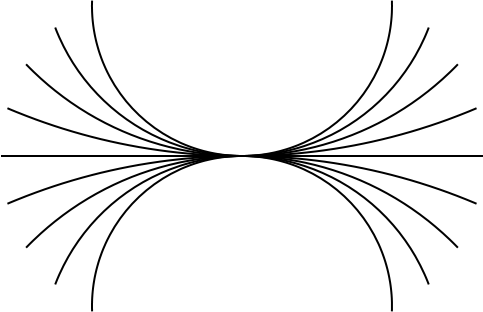} \hspace{0.5cm}
\includegraphics[height=3cm]{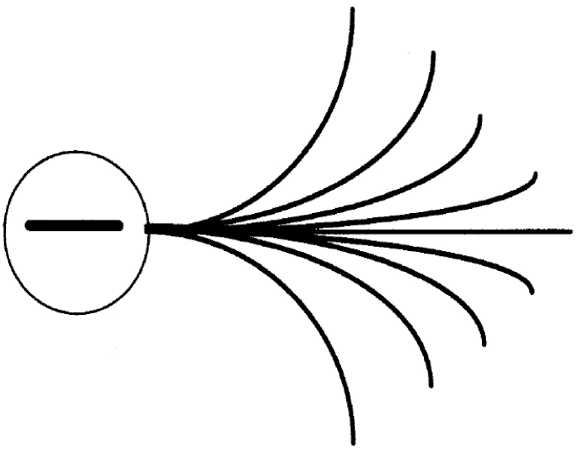}
\label{fig3}
}\\
\subfigure[ ]{
\includegraphics[width=8cm]{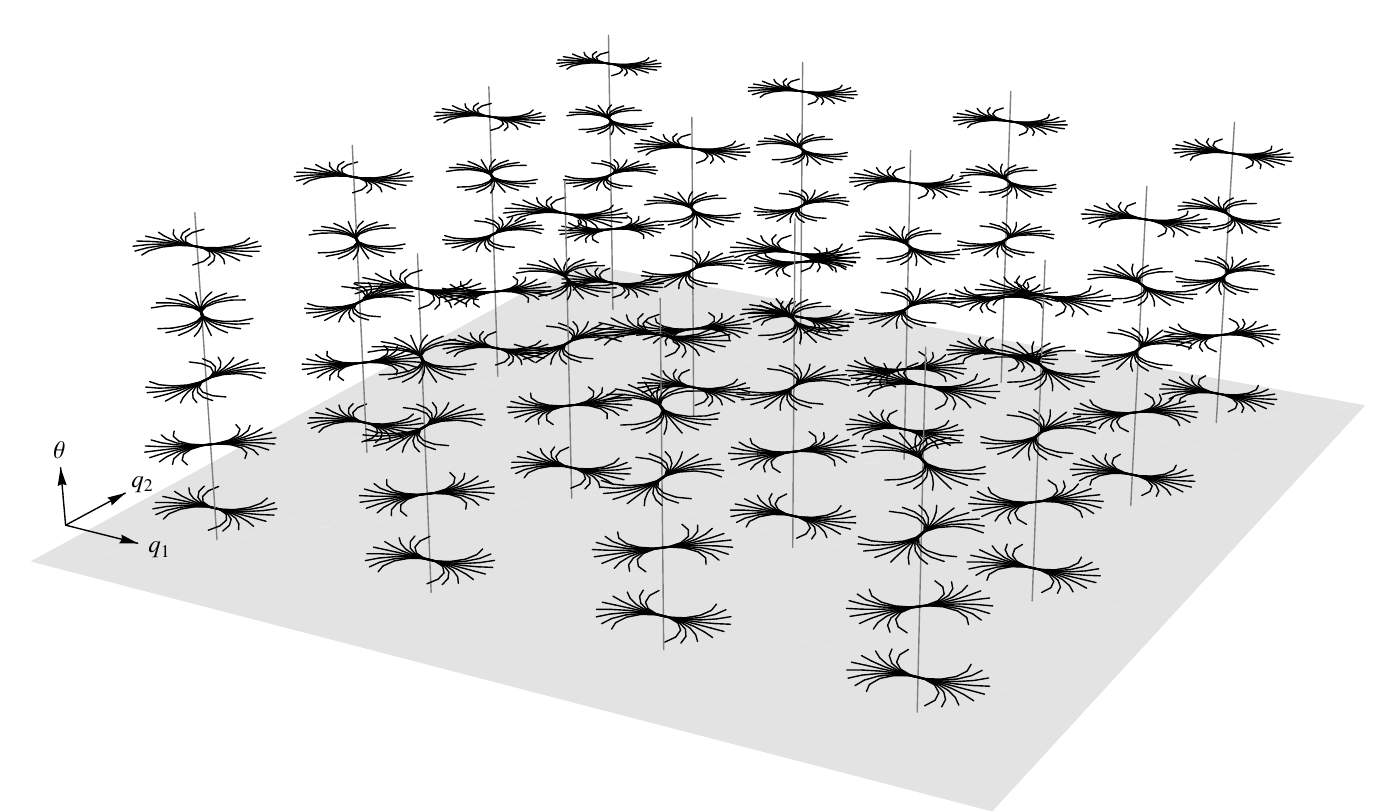}
\label{fig16}
}

\caption[ ]{{\bf{Functional architecture of V1.}} \subref{fig1} Set of receptive profiles of simple cells with constant
frequency modulus $|p|$as elements of the rotation translation group
indexed by the triple $g=(q_1, q_2, \theta)$.The parameters $(q_1,
q_2)$ code the cell position in the cortical layer and  $\theta$ is
the engrafted variable of orientation preference. The entire set of
$(q_1, q_2, \theta)$ constitutes the reduced phase space where the
frequency modulus $|p|$ is missed. \subref{fig3} A fan of integral curves of the vector fields $X_1+k X_2$ by varying the parameter $k$ (left). The curves model the connectivity between cells on the cortical plane and are implemented by neural horizontal connectivity. The model is in good agreement with the association fields of Field, Hayes and Hess \cite{Field1993Contour} (right). \subref{fig16} The entire set of fan modeling the functional architecture is obtained by applying the group action to the fan \subref{fig3}.}
\end{figure}

\subsection{Horizontal connectivity and integral curves in $SE(2)$}

Simple cells are connected by means of the horizontal connectivity
with its typical anisotropic pattern described in detail in a number of experiments (e.g. \cite{Bosking1997Orientation}).
It has been shown that the functional architecture of
cortico-cortical connections among simple cells strongly depends on position and
orientation preference, while it presents a weaker dependence on the frequency.
Indeed, while studying the connectivity,
very often in literature
only the three parameters
 $(q_1, q_2, \theta)$ are considered, while the modulus of frequency
 $|p|$ is discarded \cite{BenShahar2004Geometrical}\cite{Citti2006Cortical}\cite{Zweck2004Euclidean}\cite{Duits2007Image}.
 Moreover, the
functional architecture is invariant for change of polarity of
simple cells so that the variable $\theta$ is defined on $[0, \pi)=P^1$,
instead of $[0, 2\pi)$. Hence the polar coordinates will be expressed as $(p_1,p_2)=|p|(-\sin(2\theta), \cos(2\theta))$.
 These parameters  are the coordinates of the reduced cotangent space
 $T_R^*(C)$
which can be identified with the elements of the group
 $SE(2)$ of translations and rotations of $\R^2$, representing the
 group of symmetries of the functional architecture of V1.
 
 The group acts on a vector $(x_1,x_2,\varphi)$ applied in the origin by rotating it of an angle $\theta$ and by translating its application point of $(q_1,q_2)$:
\begin{equation} \label{rototrasla}
(x_1,x_2) \longmapsto L_{(q_1,q_2, \theta)}(x_1,x_2)=r_{2\theta}\binom{x_1}{x_2}+\binom{q_1}{q_2}\, \hbox{ and } \, \varphi \longmapsto (\varphi - \theta)
\end{equation}
 where
$$r_{2\theta}=\binom{\cos (2\theta) \ - \sin(2\theta)}{\sin (2\theta) \ \cos (2\theta)}.$$
In particular $L_{(q_1,q_2, \theta)}$ is the 2D projection of the group action.

 The tangent space of $SE(2)$ has dimension three and it is generated by the infinitesimal transformations along two orthogonal translations and the rotation of the group. However, in order to take into account the anisotropy
of the connectivity, in \cite{Citti2006Cortical} and \cite{Petitot1999Vers} just two generators instead of three were considered. In particular in \cite{Citti2006Cortical} it was proposed to
choose as generators the differential operators $X_1$ and $X_2$ representing only one infinitesimal translation 
and the infinitesimal rotation. Projecting them on the 2D cortical layer 
through the local action $L$
defined in (\ref{rototrasla}) we obtain
$$
X_1f(x_1, x_2) = \left. \frac{d}{d q_1}  \right|_{0} f( L_{(q_1,0,0)}(x_1,x_2))=
\partial_{x_1}f(x_1, x_2) $$
and $$X_2f(x_1, x_2)= \left. \frac{d}{d \theta} \right|_0 f(
L_{(0,0, \theta)}(x_1,x_2))=(x_1\partial_{x_2}-x_2\partial_{x_1})f(x_1,
x_2).
$$
These derivations are left invariant with respect to the
action $L$ and are simply the directional derivatives in the
direction respectively
 $\vec{X}_1 =(1,0)^T$ and  $\vec{X}_2=(x_2, -x_1)^T$.
The horizontal connectivity is then modeled as the set of integral
curves $\gamma(s)=(x_1(s), x_2(s))$ of the vector fields $\vec{X}_1
+ k \vec{X}_2$ having starting point at the origin of
the local coordinates and varying the parameter $k$ in $\R$ (see Fig. \ref{fig3}):
$$\gamma'(s) =(\vec{X}_1 + k \vec{X}_2)(\gamma(s)), \,\,\, \gamma(0) = 0.$$
Computing the left
invariant derivatives at the point $0$ amounts to choose local
coordinates around a fixed point. This is why the fan is
centered in the origin and oriented along the $x_1$-axis in local
 coordinates. However this choice is not restrictive, since we
 can always apply the action $  L_{(q_1,q_2,\theta)}(x_1(s), x_2(s)),$
and recover global coordinates, in which the fan has an arbitrary
origin $q$
 and direction $\theta$. Hence, in the sequel, we will
always make this choice of local coordinates.

 The cortical activity  $f=f(x_1, x_2)$
defined on the 2D cortical layer is propagated along the
horizontal long range connectivity, modeled by the $\gamma$-curves so that
the gradient of the cortical activity will be computed as directional
derivative of $f$ along the vector fields $\vec X_1$ and $\vec X_2$.

\subsection{The operators in the reduced Fourier domain}\label{sectvfields}

\begin{figure}[!ht]
\centering
\subfigure[ ]{
\includegraphics[width=8cm]{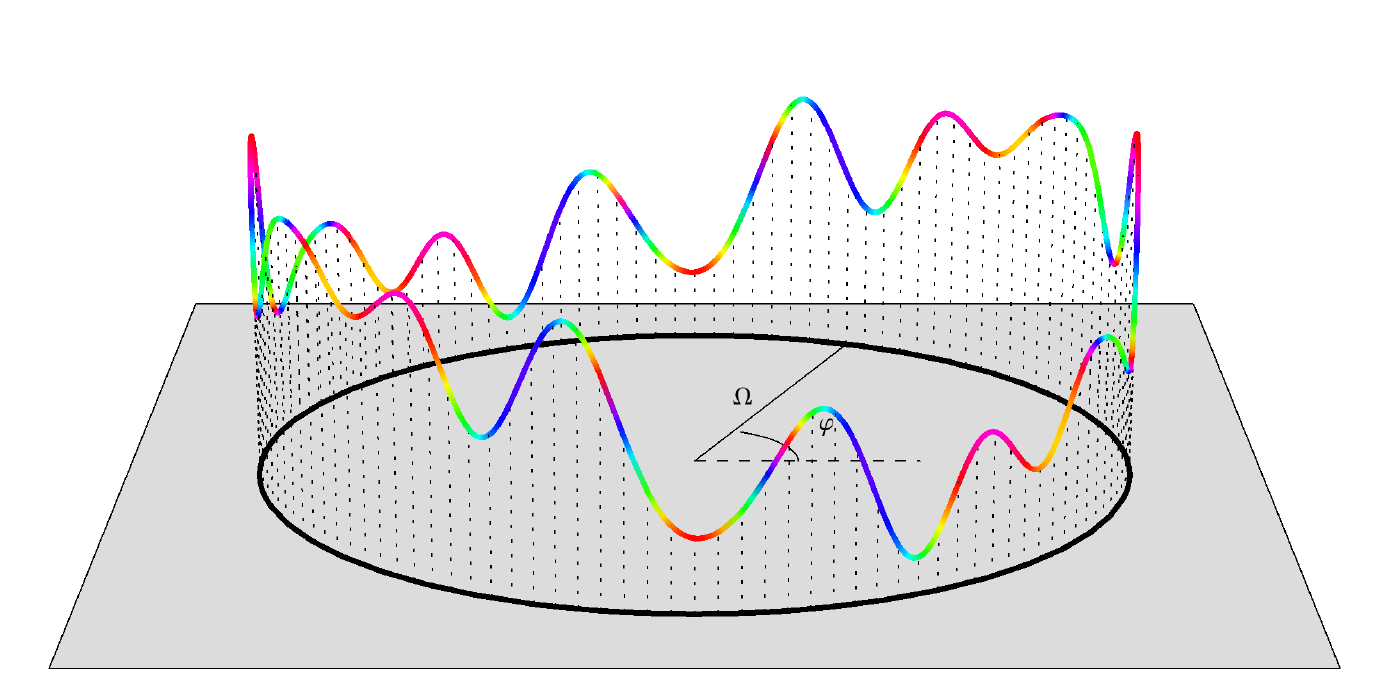}
\label{fig11}
}\subfigure[ ]{
\includegraphics[width=8cm]{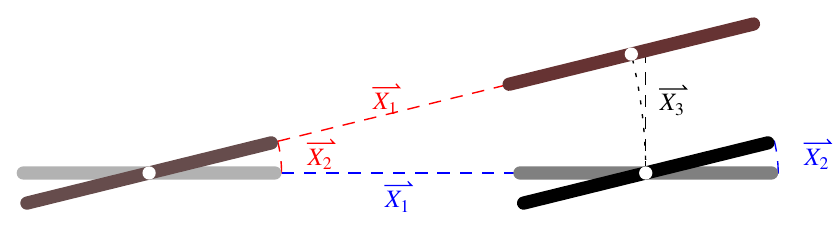}
\label{fig6}
}
\caption[ ]{ \subref{fig11}A general function $\hat u(\varphi)$ in the reduced Fourier plane, where only a circle of radius $\Omega$ is considered parametrized by the angular variable $\varphi$. The circle is a doubling  of the vertical $[0,\pi]$ segments in fig.\ref{fig1}.  $\hat u(\varphi)$ is defined on this circle with period $\pi$ and then its graph is repeated twice. The height of the graph represents $|\hat u(\varphi)|$ and the color $\arg (\hat u(\varphi)).$ \subref{fig6} The noncommutativity of the group is visualized observing that the final point achieved by a translation $X_1$ followed by a rotation $X_2$ is different from the one constructed by a rotation followed by a translation.}
\end{figure}

The reduced phase space $T_R^*(C)$ contains all the position
variables $(x_1, x_2),$ but only the angular component $\varphi\in
P^1$ of the frequency variable. Hence in the Fourier space it
supports only functions defined on a semicircle of radius
$\Omega$ fixed,  and periodical of period $\pi$ as depicted in Fig \ref{fig11}.
The operators acting on these functions are the Fourier transforms
of $X_1$ and $X_2$:
\begin{equation}\label{fourierfields}\hat X_1=-\Omega \sin(2\varphi), \quad \hat X_2 =
i \partial_{\varphi} .\end{equation} Let us note that the fields
$\hat X_1$ and $\hat X_2$ correspond respectively to angular
position and angular momentum of the $\pi$ periodic quantum pendulum \cite{Cohen1978Quantum}. The parameter $\Omega$ is specified by the physics of the
problem and in our case it will express the reciprocal of the correlation length of pinwheels in accordance with \cite{Niebur1994Design}. 

\section{An uncertainty principle on the functional geometry}
\subsection{The noncommutativity of vector fields}

While studying the
relation between angular position and momentum, both in the real and Fourier plane,
we have to take into account that the composition of a translation and a rotation is not commutative: $$r_{2\theta}\binom{x_1}{x_2} +\binom{q_1}{q_2} \not= r_{2\theta}\left(\binom{x_1}{x_2} +\binom{q_1}{q_2}\right).$$  
The difference for
$\theta$ sufficiently small is an increment in the direction
orthogonal to $q$ (see Fig. \ref{fig6}). This property implies the non
commutativity of left invariant derivatives $X_1$ and $X_2$. In
particular if we compute second mixed derivatives we have that $
 X_2 X_1 f(x_1, x_2) \not= X_1 X_2 f(x_1, x_2).
$ We will call commutator of $X_1$ and $X_2$ the difference
$X_3f =  X_2 X_1 f - X_1 X_2 f,$
which is a measure of the non commutativity of the space.
The same property can
be transferred in the transformed fields $\hat X_1$ and $\hat X_2$ whose commutator is
$$\hat X_3   =  \frac{i}{2}\left(\hat X_2 \hat X_1  - \hat X_1 \hat X_2\right)  = \Omega \cos(2\varphi).$$

\subsection{The uncertainty principle}

This noncommutative framework has important analogies with the
canonical quantum mechanical case, in which the two non commuting operators are
position and momentum. In that specific situation the variance of the
position and momentum operators on functions have a lower bound that
has been formalized in the well known Heisenberg uncertainty
principle. This principle has been formally extended (see
\cite{Folland1989Harmonic} and references therein) to general non commuting vector fields.
Hence we can express here a similar uncertainty principle in terms
of the vector fields  $\hat X_1$ and $\hat X_2$
 \begin{equation} \label{principle}
\Delta_{\hat f} \hat X_1 \, \Delta_{\hat f} \hat X_2 \geq \frac12 \left|<\hat X_3 >_{\hat f}\right|,
 \end{equation}
where $\Delta_{\hat f} \hat X$ is the standard deviation of the operator $\hat X$ and $|<\hat X >_{\hat f}|$ is the mean value of the operator $\hat X$ when evaluated on the function $\hat f$. For  $|<\hat X_3 >_{\hat f}|$ fixed we recognize that,
if $\hat f$ is concentrated in the Fourier domain so that its variance is small, then the variance of the angular
momentum has to be high and viceversa.
Note that the inequality expresses the uncertainty principle for the $\pi $ periodic quantum
pendulum between the two observable angular position and angular
momentum. For $\varphi \rightarrow 0$, the fields $\hat X_1$ and
$\hat X_2$ reduce to the classical position and momentum operators
of the harmonic oscillator and the inequality expresses the well
known Heisenberg uncertainty principle.

\section{Coherent states as optimal localization in angular position and angular momentum}
\subsection{Coherent states in the Fourier space}

The minimizers of the uncertainty principle are called coherent states and correspond to the functions with the optimal localization in angular position and angular momentum.
Let us first compute the reference minimal uncertainty ``ground''
state $\hat u_0^\Omega(\varphi)$ by making use of the well-known equation
(see \cite{Folland1989Harmonic})
\begin{equation} \label{Follandeq}
\hat X_2 \hat u_0^\Omega (\varphi) =2 i \lambda \hat X_1 \hat u_0^\Omega(\varphi),
\end{equation}
where the scaling parameter $\lambda$ represents the frequency of the pendulum.
Its solution can be analytically computed and, up to a normalization constant, is the ground state
$\hat u_0^\Omega(\varphi) =  e^{\lambda\Omega\cos(2\varphi)}$.
The set of coherent states of $SE(2)$ in the
Fourier domain depends on the angular parameter
$\theta$ and is obtained as a displacement of the ground state $\hat u_0 (\varphi)$  of an angle $\theta$:
\begin{equation}\label{fouriercoherent}
\hat u_\theta^\Omega(\varphi) =  e^{\lambda\Omega\cos(2(
\varphi- \theta))} .
\end{equation}

\begin{figure}[!ht]
\centering
\subfigure[ ]{
\includegraphics[width=7cm]{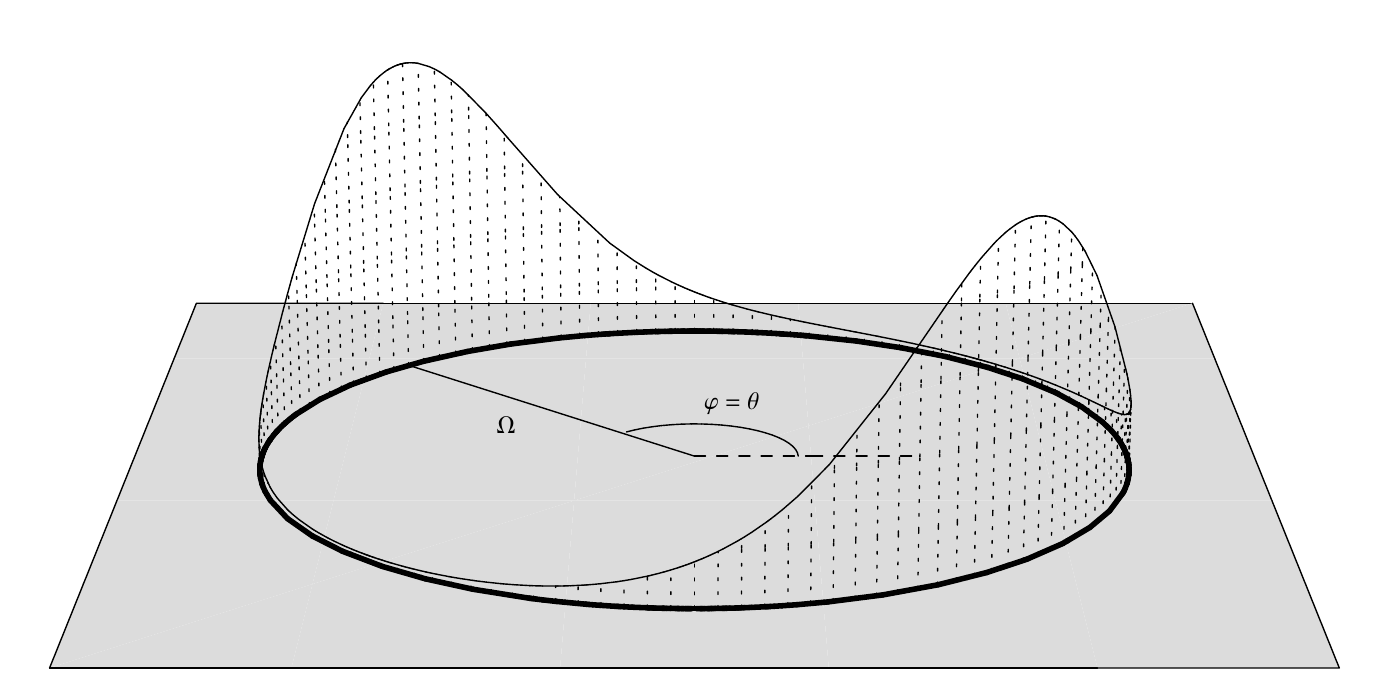}
\label{fig7}
}\subfigure[ ]{
\includegraphics[width=8cm]{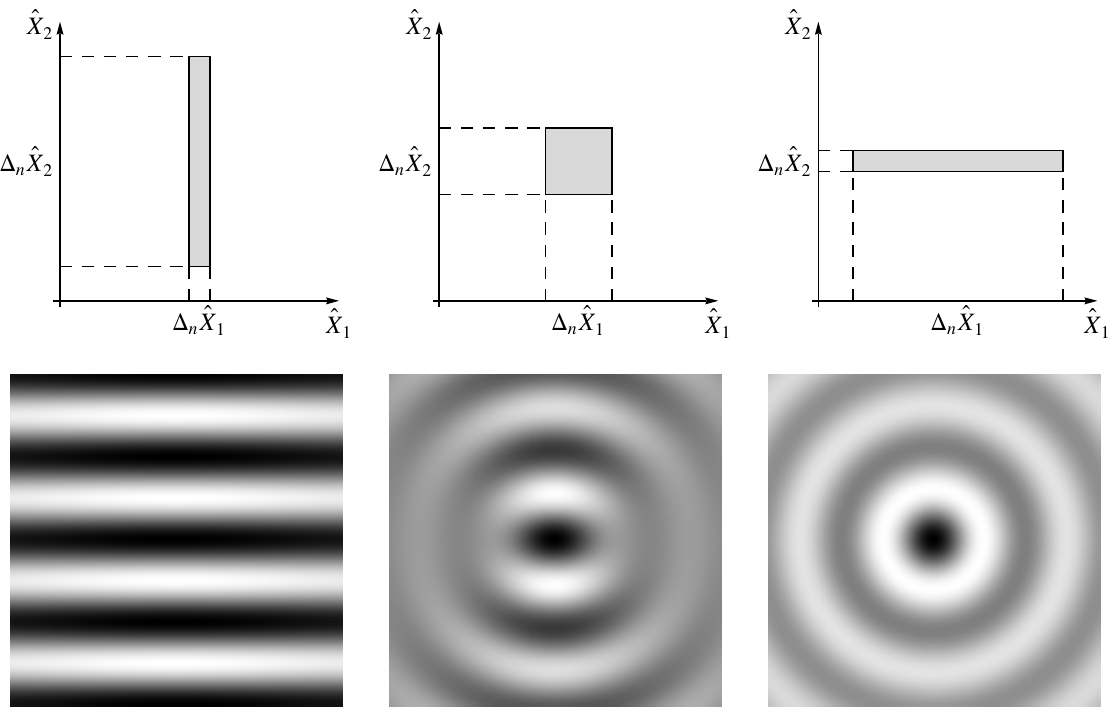}
\label{fig10}
}\\
\subfigure[ ]{
\includegraphics[width=7cm]{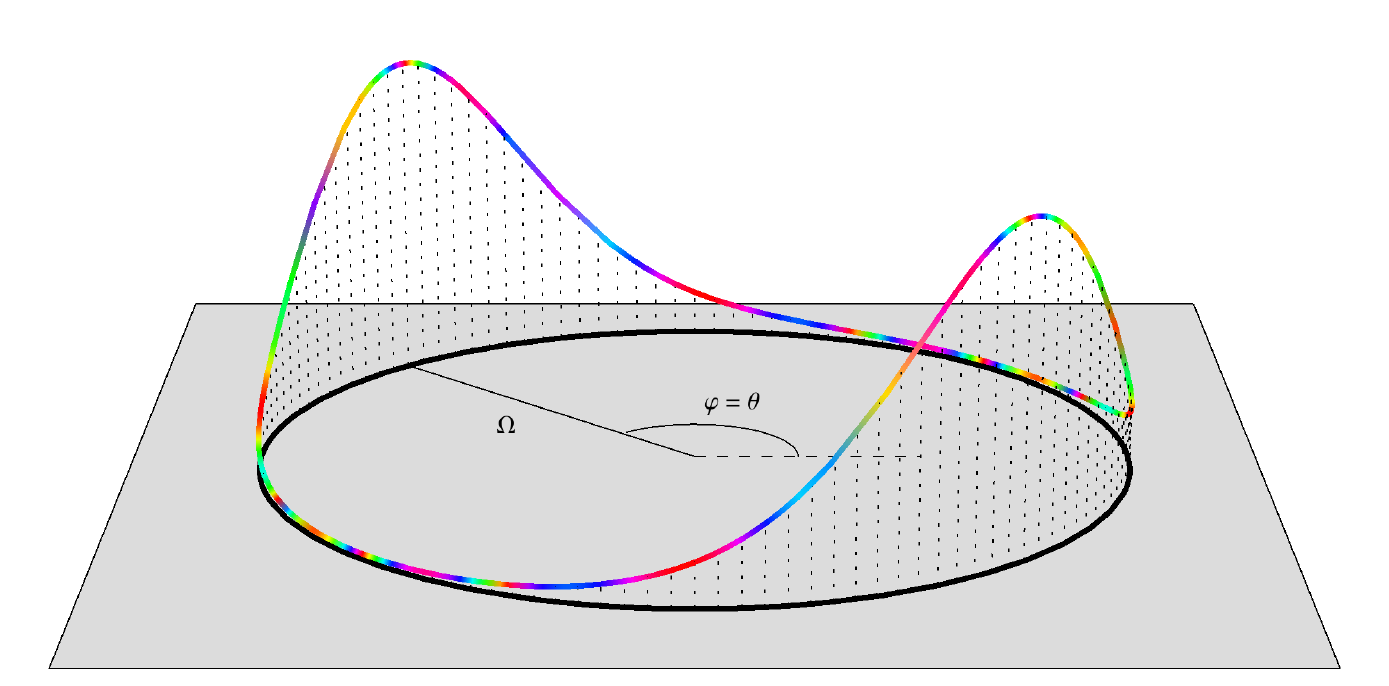}
\label{fig8} \hspace{.5cm}
}\subfigure[ ]{
\includegraphics[width=3.5cm]{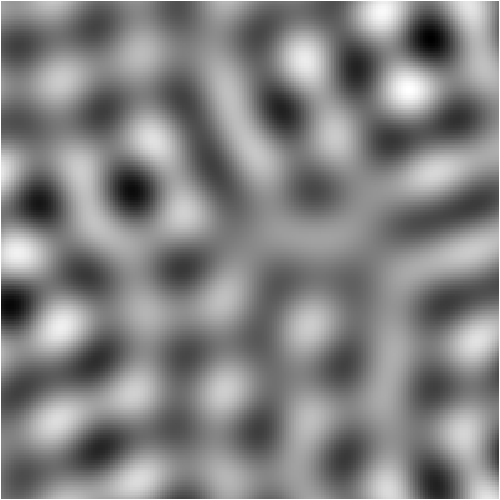}
\includegraphics[width=3.5cm]{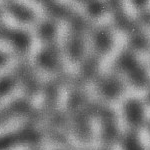}
\label{fig12} \hspace{.5cm}
}
\caption[ ]{\subref{fig7} A coherent state $\hat u_\theta^\Omega(\varphi)$ in  the reduced Fourier plane for a fixed value of $\theta$ that is a real valued function minimizing the uncertainty principle (\ref{principle}). \subref{fig10} (top) The standard deviation of the angular position $\Delta_{\hat u^\Omega_\theta}\hat X_1$ versus the one of angular momentum $\Delta_{\hat u^\Omega_\theta} \hat X_2$ for three different values of $\lambda$: $\lambda \gg \lambda = \frac{1}{2\Omega}$ (left), $\lambda = \frac{1}{2\Omega}$ (center), $\lambda \ll \frac{1}{2\Omega}$ (right). For a better visualization the standard deviation is normalized $\Delta_n X_i = \Delta_{\hat f} \hat X_i / |\langle \hat X_3 \rangle_{\hat f}|^{\frac12},$ $i=1,2$. 
(bottom) Coherent states in the real plane obtained by anti-transforming $\hat u_\theta^\Omega(\varphi)$.  For $\lambda \rightarrow +\infty$ the states $u$
is maximally concentrated in the angular position (orientation) and
it results $u(x_1,x_2)= e^{i \Omega (-x_1 \sin(2\theta) + x_2
\cos(2\theta))}$ that is a pure harmonic function, whose level sets
are rectilinear and globally oriented (left). Note that in this case the $X_1$ derivative in global coordinates vanishes i.e. $X_1 u(x_1,x_2)=0$ meaning that the angular momentum is maximally spread
and the angular position is maximally localized. In the other case
$\lambda \rightarrow 0$ the state becomes a Bessel function
$u(x_1,x_2)=\int_0^\pi e^{i\Omega\sqrt{x_1^2 + x_2^2}\cos(2(\varphi
- \theta))}d\varphi$  achieving the best concentration
in angular momentum, since its level sets are circles with all the
possible curvatures (right). Note that in this case $X_2 u(x_1,x_2)=0$
meaning that the angular position is maximally spread and the
angular momentum is maximally localized. For $\lambda$ between the
two limiting conditions the uncertainty are distributed between
angular position and  momentum achieving the equal distribution of
uncertainty  for $\lambda = \frac{1}{2\Omega}$ (center). \subref{fig8} A covariant coherent state $\hat u_\theta^\Omega(\varphi)e^{i \alpha(\varphi)}$ in the reduced Fourier plane for a fixed value of $\theta$, solution of the equation (\ref{Follandeq}) with covariant derivatives. The set of states is invariant with respect to local change of coordinates both in the domain and co-domain modeling a system that does not require any knowledge of an external global reference. The function is visualized with the same coding as in figure \ref{fig11}. \subref{fig12} The Fourier anti-transform of a covariant coherent state (left) as model of a real activity map recorded by Bosking et Al \cite{Bosking1997Orientation} (right).}
\end{figure}

Every coherent state represents the most concentrated function
for optimal localization in angular position and momentum, both
taken starting from the reference angle $\theta$. The function $\hat u_\theta^\Omega(\varphi)$
 is bell-shaped and centered around the angle
$\theta$ (see Fig. \ref{fig7}). We note that the bigger is $\lambda$, the sharper is its
localization. Indeed its standard deviation can be evaluated as $
\Delta \varphi \approx \frac{1}{2\sqrt{\Omega \lambda}} $ and its width
reduces around $\varphi = \theta$ as $\Omega \lambda$ becomes big,
leading to a delta $\delta(2(\varphi-\theta))$ in the limit.

If $\hat u = \delta(\varphi)$ then $\sin(\varphi) \delta(\varphi) = 0$ and the variance of the angular position is null, meaning that it is maximally concentrated. Due to the uncertainty inequality $\partial_\theta \delta(\varphi)$ has infinite variance i.e. it is maximally undetermined. On the other hand, if $\hat u$ is constant, then $\partial_\theta \hat u = 0$ and the angular momentum is maximally concentrated.

Up to now, the coherent states $\hat u_\theta^\Omega(\varphi)$ have
been computed using left invariant operators of the group $SE(2)$ and hence imposing invariance on the domain of functions.
Now we impose invariance also on the co-domain of $\hat
u_\theta^\Omega(\varphi)$ with respect to local phase rotation $e^{i
\alpha(\varphi)}$ so that  each function  $\hat
u_\theta^\Omega(\varphi)e^{i \alpha(\varphi)}$ is a coherent state.
These states are solutions of the
 equation (\ref{Follandeq}) where the operators  $\hat X_1$ and  $\hat X_2$ are
expressed  in terms of covariant derivatives instead of the usual
derivatives. {\it These coherent states, visualize in Fig \ref{fig8} are invariant for any local
change of coordinates in the domain and co-domain, modeling a system
that does not require any knowledge of an external global reference
system.}

\subsection{Coherent states in the real plane}

The Fourier anti-transform of the function $\hat
u_\theta^\Omega(\varphi)e^{i \alpha(\varphi)}$ defined on a circle
of radius $\Omega$ gives the expression $u_{\theta}(x_1,x_2)$ of the
coherent states in the real plane, providing the maps visualized in
figure.

\subsection{From coherent states to pinwheels}

\begin{figure}[!ht] \label{fig9}
\centering
\includegraphics[width=8cm]{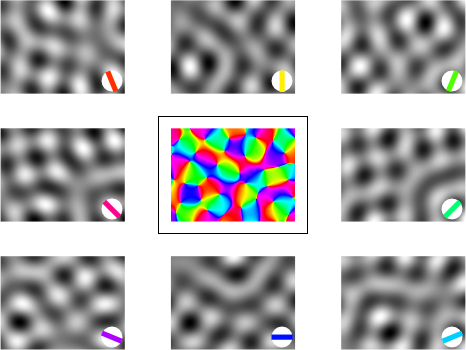} \hspace{1cm}
\caption{Gray valued images visualize covariant coherent states on the 2D real plane by varying the angle $\theta$. The color image in the center has
been constructed from gray valued maps, associating a color
coding representation to preferred orientations, as in  \cite{Bonhoeffer1991Isoorientation}.}
\end{figure}

In \cite{Bonhoeffer1991Isoorientation} the pinwheel structure of V1 has been reconstructed starting from a set of
cortical activity maps acquired with
optical imaging techniques in response to gratings with different
orientations $\theta$. A color image has
been obtained from gray valued activity maps, associating a color
coding representation to preferred orientations \cite{Swindale1987Surface}. Here we apply the
same procedure to the set of coherent states (our model of activity maps), producing the pinwheel
structure visualized at the center of Fig. \ref{fig9}.

\section{Qualitative and quantitative comparison with  orientation cortical maps}
The image show a qualitative impressive similarity with the original result obtained in \cite{Bonhoeffer1991Isoorientation} \cite{Blasdel1991Orientation} \cite{Bosking1997Orientation} among many others. Moreover by computing the power spectrum of recorded pinwheels, following \cite{Niebur1994Design},  a function defined on an annulus with radius $\Omega$ is obtained, in complete accordance with the result predicted by the presented theory, built up on the basis of the geometry of the functional architecture. In our perspective the orientation  activity maps are modeled by the coherent states, minimizing uncertainty introduced by functional geometry. Note that the coherent states are $\pi$ periodic as the measured activity maps.
To our knowledge in the existing models of pinwheels the activity maps are obtained a posteriori, while in our approach a geometric model of the activity maps is directly provided and the pinwheels are constructed as in the classical experiments.

\newpage

\bibliographystyle{unsrt}
\bibliography{paperpin}

\end{document}